\documentclass[twocolumns,showpacs,preprintnumbers,amsmath,amssymb]{revtex4}
\usepackage{graphicx}

\begin{document}

\title{Fermi-Bose quantum degenerate $^{40}$K-$^{87}$Rb mixture with attractive
interaction}

\author{G. Roati}
\altaffiliation[Also at: ]{Dipartimento di Fisica, Universit\`a di
    Trento, 38050 Povo, Italy.}
\author{F. Riboli}
\author{G. Modugno}
\author{M. Inguscio}
 \affiliation{ European Laboratory for Nonlinear Spectroscopy and Dipartimento di Fisica, Universit\`a di Firenze, \\
 and INFM, Via Nello Carrara 1, 50019 Sesto Fiorentino, Italy }

\date{\today} 
\begin{abstract}
We report on the achievement of simultaneous quantum degeneracy in
a mixed gas of fermionic  $^{40}$K and bosonic $^{87}$Rb.
Potassium is cooled to 0.3 times the Fermi temperature by means of
an efficient thermalization with evaporatively cooled rubidium.
Direct measurement of the collisional cross-section confirms a
large interspecies attraction. This interaction is shown to affect
the expansion of the Bose-Einstein condensate released form the
magnetic trap, where it is immersed in the Fermi sea.
\end{abstract}

\maketitle

The recently demonstrated quantum degeneracy of Fermi-Bose (FB)
mixtures of dilute atomic gases \cite{hulet,salomon,ketterle1}
promises to further enrich the field of the physics of degenerate
matter at ultralow temperatures \cite{ketterle,varenna}. When a
Bose-Einstein condensate (BEC) interacts with a Fermi gas, novel
phenomena are expected to occur. The most appealing one is
certainly BCS-like fermionic superfluidity, since a BEC could
affect interactions between fermions \cite{bijlsma,viverit}.
Furthermore, different FB interaction regimes could allow studies
of phase-separation \cite{molmer,tosi} or of the stability
properties of the binary mixtures \cite{roth1}. \par The mixtures
so far reported have in common the use of fermionic $^{6}$Li,
combined with a BEC of $^{7}$Li \cite{hulet,salomon}, or of
$^{23}$Na \cite{ketterle1}. For the $^{6}$Li-$^{7}$Li mixtures the
FB interaction is \emph{repulsive}, with possible consequences for
the separation of the components, and eventually for the thermal
contact. For the $^{6}$Li-$^{23}$Na case, the interaction has not
been measured, however the theoretical predictions are again in
favor of a repulsive character \cite{timmcote}.\par A different,
promising scenario would be offered by mixtures combining species
with attractive interaction, because the absence of a phase
separation would allow efficient cooling well below the Fermi
temperature and would favor the interaction between the two
components in the degenerate regime. With this respect, a mixture
composed of $^{40}$K and $^{87}$Rb would be particularly
interesting. Indeed, precise $^{41}$K-$^{87}$Rb interspecies
collisional studies at ultralow temperatures \cite{nostrocoll},
inferred an \emph{attractive} character for the interaction of the
$^{40}$K-$^{87}$Rb pair. \par In this Letter we report the
production of such a novel macroscopic quantum system, in which a
degenerate $^{40}$K Fermi gas, composed of more than 10$^4$ atoms,
coexists with a $^{87}$Rb Bose-Einstein condensate of up to
2$\times$10$^4$ atoms. The large interspecies scattering length
results in an efficient sympathetic cooling of $^{40}$K with
evaporatively cooled Rb, as in the case of Bose-Einstein
condensation of $^{41}$K \cite{science}. This cooling scheme
represents also an alternative to the single-species evaporation
approach \cite{jin} that was early demonstrated to produce a Fermi
gas of K atoms. We observe signatures of the large interaction of
the two components also in the degenerate regime.\par The
degenerate mixture is produced using the apparatus described in
Ref.\cite{science}. In brief, about 10$^{5}$ $^{40}$K atoms and
5$\times$10$^{8}$ $^{87}$Rb atoms at a temperature around
100~$\mu$K are loaded in an elongated magnetostatic trap using a
double magneto-optical trap apparatus. As opposed to the case of
$^{41}$K \cite{science}, combined magneto-optical trapping of
$^{40}$K and $^{87}$Rb is efficient, as was also shown in
Ref.~\cite{jin1}. Prior to magnetic trapping, both species are
prepared in their doubly polarized spin state, $|F=9/2,
m_F=9/2\rangle$ for K and $|2, 2\rangle$ for Rb. These states
experience the same trapping potential, with axial and radial
harmonic frequencies $\omega_{a}=2\pi \times 24$~s$^{-1}$ and
$\omega_{r}=2\pi\times 317$~s$^{-1}$ for K, while those for Rb are
a factor $(M_{Rb}/M_K)^{1/2}\approx 1.47$ smaller. Evaporative
cooling is then performed selectively on the Rb sample. Due to the
different gyromagnetic factors of the two species a
radio-frequency evaporation scheme could be implemented
\cite{evaporazione}, in contrast to the boson-boson mixture, for
which microwave radiation had to be used.\par With an evaporation
ramp lasting about 25~s we are able to cool typically
2$\times$10$^{4}$ K atoms and 10$^{5}$ Rb atoms to below 1~$\mu$K.
Sympathetic cooling of $^{40}$K with $^{87}$Rb is very efficient,
with a large ratio of "good" elastic collisions to "bad" inelastic
collisions. We have measured the interspecies scattering
cross-section by performing a rethermalization measurement at a
temperature around 400~nK, at which both species are still
non-degenerate gases. We drive the mixture out of equilibrium with
a short, Rb-selective, parametric heating phase \cite{nostrocoll},
and we observe the subsequent heating of K, which is mediated by
elastic interspecies collisions. For a K sample composed by
1.2$\times 10^4$ atoms, coexisting with 4$\times 10^4$ Rb atoms we
measure a short thermalization time $\tau$=57(20)~ms. At these
ultralow temperatures the collisions have almost exclusively a
$s$-wave character and, following the model discussed in
\cite{nostrocoll} $\tau$ is linked to the scattering length $a$
through $\tau^{-1}=4 \pi a^2 \xi \bar{n} v / \alpha_s$, where
$\bar{n}=(\frac{1}{N_K}+\frac{1}{N_{Rb}})\int n_K n_{Rb} d^3x $ is
the effective density of K and Rb atoms, $v$ is their relative
velocity, $\xi \approx 0.86$ is a factor which takes into account
the different mass of two colliding atoms, and $\alpha_s \approx
2.7$ is the average number of collisions needed for
thermalization. Even thought K atoms cannot thermalize between
themselves, since $s$-wave collisions are forbidden for identical
fermions, the thermalization with Rb happens on a timescale longer
than the mean period of oscillation in the trap, and this ensures
thermalization of the K sample. Actually we have observed that
both the density and momentum distributions have gaussian profiles
that lead to the same temperature. From the measured $\tau$ we
derive a quite large magnitude for the $^{40}$K-$^{87}$Rb
scattering length: $|a|$=330$^{+160}_{-100}$~$a_0$. Here the
uncertainty is dominated by that on $\tau$ and by a 40\%
uncertainty on the atom number. The direct measurement with the
fermionic isotope is in agreement with the value
$a$=-261$^{+170}_{-159}$~$a_0$ that we previously inferred by
mass-scaling from collisional measurements on the bosonic
$^{41}$K-$^{87}$Rb mixture \cite{nostrocoll}. The measurement of a
large value for $|a|$ also confirms the {\it attractive} character
of the interaction, since a positive scattering length would have
been compatible only with a much smaller magnitude
\cite{science}.\par

\begin{figure}
\begin{center}
\leavevmode
\centerline{\includegraphics[width=16cm,clip=]{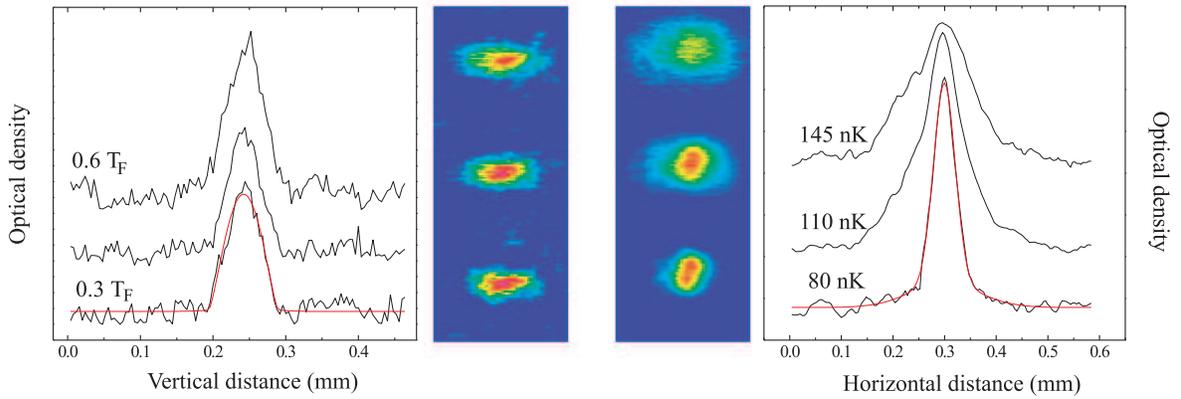}}
\end{center}
\caption{Simultaneous onset of Fermi degeneracy for $^{40}$K
(left) and of Bose- Einstein condensation for $^{87}$Rb (right).
The absorption images are taken for three decreasing temperatures,
after 4.5~ms of expansion for K and 17.5~ms for Rb, and the
sections show the profile of the momentum distributions. The
bosons provide a thermometry of the system: in the coldest sample,
the Rb BEC has a 40\% thermal component, and the temperature is
$T$=0.74$T_c$=80~nK, which corresponds to $T$=0.3$T_F$ for K. The
Thomas-Fermi radius of the K profile is $R$=52~$\mu$m, consistent
with the radius expected for a degenerate Fermi gas. }
\label{fig1}
\end{figure}

By further cooling the mixture we have evidence of the formation
of a degenerate Fermi gas coexisting with a Bose-Einstein
condensate. In Fig.~\ref{fig1} we show a series of absorption
images of the mixture for three different final energies of the
evaporation ramp, which reveals the different nature of the two
degenerate gases. Both samples are imaged in the same experimental
run, by using two short, delayed light pulses. The images are
taken after a ballistic expansion appropriate to measure the
momentum distribution of the samples; in particular the expansion
lasts 4.5~ms for K and 17.5~ms for Rb. Sections of such images are
also shown: they are taken along the vertical direction for K, and
along the horizontal direction for Rb. With our experimental
parameters, we have a Fermi temperature $T_F$=250~nK and a
critical temperature for BEC $T_c$=110~nK for a sample composed of
10$^4$ and 2$\times 10^4$ atoms, respectively. \par Thermometry of
the system is provided by the bosonic component, assuming thermal
equilibrium between the two components. As the temperature is
decreased by almost a factor of two (from top to bottom in
Fig.~\ref{fig1}), Rb undergoes the phase-transition to BEC, while
the width of the fermionic component remains almost constant. A
fit of the coldest K cloud with a Thomas-Fermi profile
\cite{butts} gives a radius $R$=52~$\mu$m, which is consistent to
within 10\% with the minimum radius allowed by Fermi statistics:
$R=R_F \sqrt{1+\omega_r^2\tau^2}$, where $R_F=\sqrt{2k_B
T_F/(M\omega_r^2)}$ is the Fermi radius and $\tau$ is the
expansion time. For the {\it quasi-pure} BEC shown in
Fig.~\ref{fig1}, a fit to both the thermal and condensate
components with gaussian profiles indicate a condensate fraction
of 60\%. This implies a temperature $T$=80~nK of both the BEC and
the Fermi gas, which corresponds to 0.3~$T_F$.\par In these
conditions, we see a small decrease of the number of K atoms on
the timescale of 1~s, which is the lifetime of the BEC in our
system. This indicates the presence of some kind of inelastic
collisional process, which will be the object of future
investigation.\par
\begin{figure}
\begin{center}
\leavevmode
\centerline{\includegraphics[width=8cm,clip=]{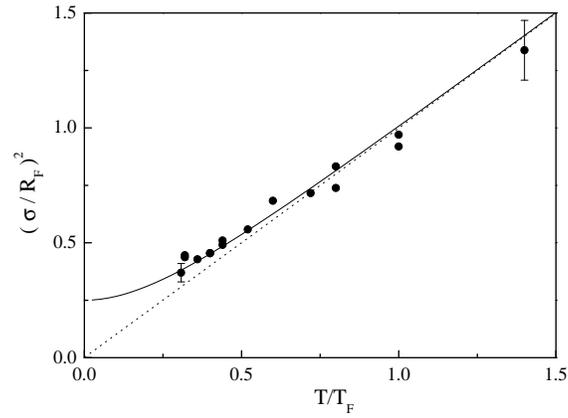}}
\end{center}
\caption{Gaussian 1/$e$ radius of the radial distribution of K atoms, versus the
reduced Fermi temperature. The temperature is given by the Rb sample and
$T_F$=250~nK. The solid line is the theoretical prediction for an ideal Fermi
gas, while the dotted line is the classical behavior.}
\label{fig2}
\end{figure}

 Further evidence for the achievement of quantum
degeneracy is obtained by studying the gaussian $1/e$ width of the
fermionic sample as a function of temperature. As shown in
Fig.~\ref{fig2} the square of the width, normalized to $R_F$,
scales linearly for $T>T_F$, indicating thermal equilibrium
between K and Rb. Below $T_F$, the data deviate from the behavior
expected for a classical gas, and indeed they are better
reproduced by the prediction of the model for an ideal Fermi gas
\cite{butts}.\par Since we have also observed degenerate mixtures
in which the thermal fraction of the condensate was below our
detection limit of nearly 30\%, the attainment of temperatures
lower than those reported cannot be excluded. However, boson
thermometry is no longer possible in this regime, and different
techniques would be necessary to investigate the evolution of
sympathetic cooling when both species are well below their
critical temperatures. \par
\begin{figure}
\begin{center}
\leavevmode
\centerline{\includegraphics[width=8cm,clip=]{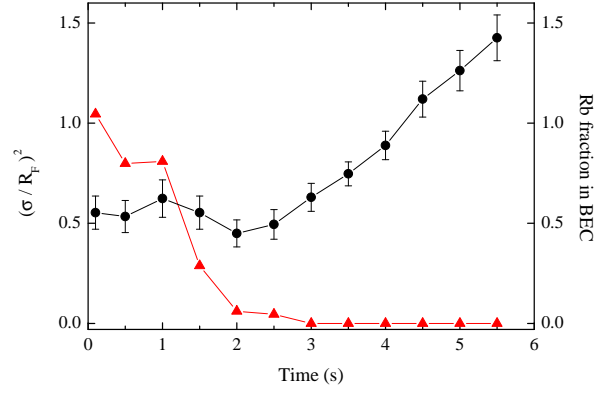}}
\end{center}
\caption{Thermal exchange between the two degenerate gases. The
gaussian width of $^{40}$K (circles) increases only when $^{87}$Rb
atoms (triangles) are almost completely evaporated from the trap,
as explained in the text. Each data point is the average of three
or four measurements, and the solid lines are a guide to the eye.}
\label{fig3}
\end{figure}

 In the present experiment we have an
evidence for thermal contact between the two species in the
degenerate regime, even when no thermal component is detectable
for the Rb BEC. This is obtained by leaving the degenerate mixture
in the magnetic trap for a relatively long time after the end of
the evaporation. The Rb temperature is kept constant by means of a
radio-frequency shield, but the background heating ($\approx$
100~nK/s) caused by fluctuations in the magnetic field,
continuously removes atoms from the BEC. This is illustrated in
Fig.~\ref{fig3}, together with the simultaneous behavior of K. The
evolution of the width of the fermionic distribution indicates
that K starts to heat up only when Rb is almost completely
evaporated from the trap. Although the gaussian width is not a
sensitive "thermometer" at low temperatures, as shown in
Fig.~\ref{fig2}, the results reported in Fig.~\ref{fig3} are
significant. Indeed, if K were thermally decoupled from Rb, its
heating at the observed rate would be detectable after 1~s even in
the extreme case of a starting temperature $T<<T_F$. \par However,
it is difficult to determine whether the thermal contact is direct
or mediated by a possible, undetected thermal cloud. In our
magnetic trap, the centers of mass of the two species are
displaced due to the different gravitational sag for K and Rb.
This displacement, $\Delta z=2.9$~$\mu$m, is not sufficiently
large to affect the geometrical overlap of the two degenerate
components, since the radial sizes of the Fermi and Bose gases are
$R_F$=5.1~$\mu$m and $R_B$=2~$\mu$m, respectively. Therefore, the
BEC is completely immersed in the Fermi sea, with a ratio of the
two volumes of approximately 1:16, hence direct thermal exchange
is possible. However, the contribution of an undetected thermal
cloud cannot be excluded, also because of the large K-Rb
interaction. For instance at $T$=0.5$T_F$ we calculate that 10$^4$
uncondensed bosons would thermalize with an equal number of
fermions in about 50~ms. \par
\begin{figure}
\begin{center}
\leavevmode
\centerline{\includegraphics[width=8cm,clip=]{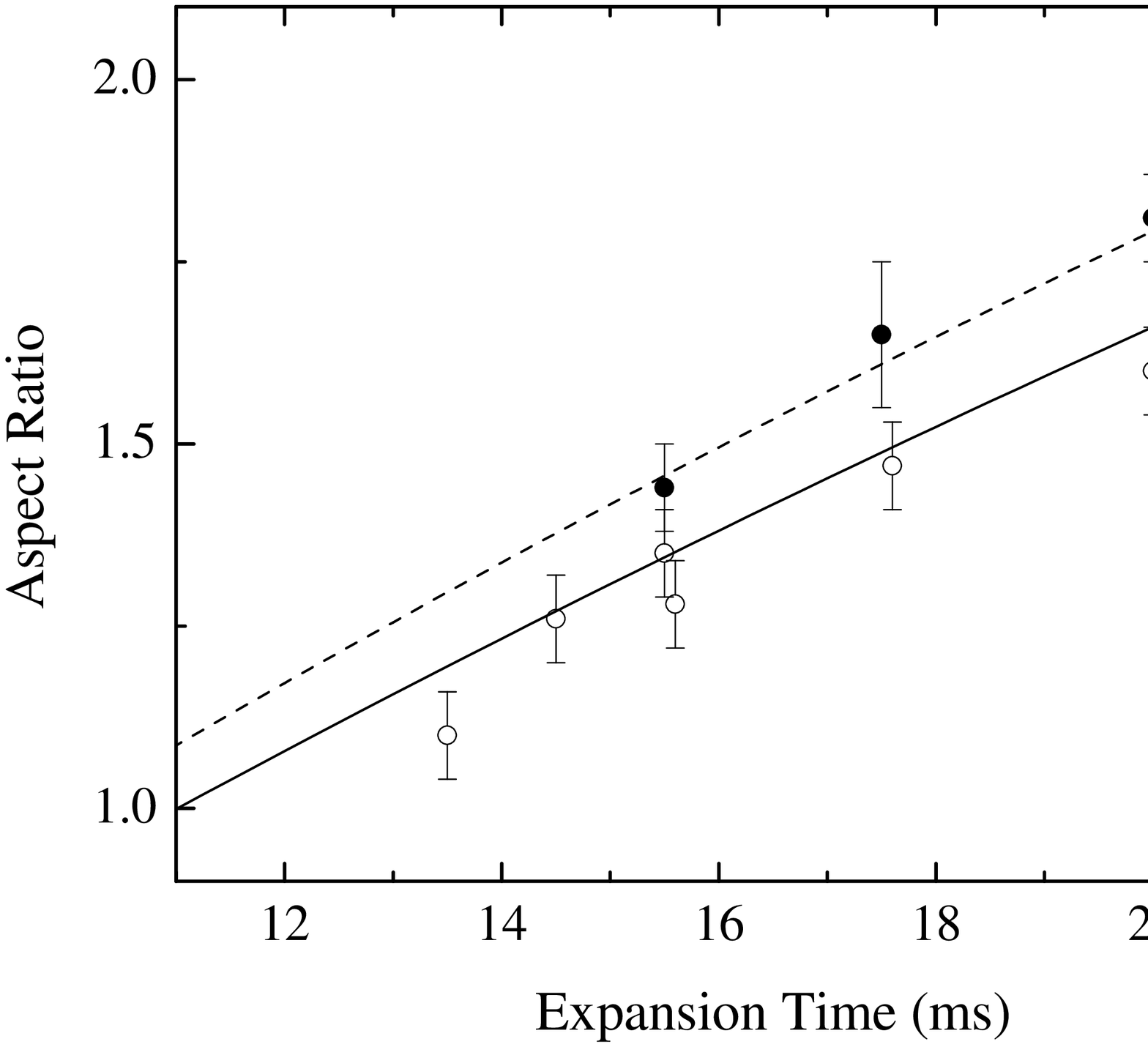}}
\end{center}
\caption{Modification of the expansion of a Rb BEC due to
interaction with the K Fermi gas. The radial-to-axial aspect ratio
increases more rapidly with time for condensates created with K
(solid circles) than for pure condensates (open circles). The two
curves are the theoretical predictions for the expansion of a pure
BEC for the trap parameters of the present experiment (solid line)
and for trap frequencies 10\% larger (dashed line).} \label{fig4}
\end{figure}

 The large attractive interspecies
interaction is expected to affect the density profile of both
degenerate gases \cite{roth1,roth}. Our experimental configuration
is more suitable for the observation of the effect of the mutual
interaction on the boson. While the Rb BEC is completely immersed
in the Fermi sea, only a relatively small volume fraction of K is
exposed to the attraction. In fact, we have seen a modification on
the ballistic expansion of the BEC due to the presence of the
other species. In general, we found that BECs coexisting with the
Fermi gas invert their aspect ratio more rapidly than normal
during the expansion. As an example, the evolution of the
axial-to-radial aspect ratio of the condensate is shown in
Fig.~\ref{fig4} for both the cases of a pure BEC and of a BEC with
the Fermi gas. In the former case the expansion is in agreement
with the theoretical prediction \cite{stringari} for the
experimental trap parameters. The latter case is instead better
reproduced by the behavior expected for a pure BEC confined in an
effective potential with frequencies 10\%larger than the actual
ones. This constitutes a direct evidence of the interaction, and
qualitatively agrees with the expectations reported in
Ref.~\cite{roth} of a tighter confinement for a BEC in a Fermi gas
with mutual attraction. However, a precise reconstruction of the
BEC density profile in the trap is not straightforward from these
observations. Indeed, in modelling the data of Fig.~\ref{fig4} as
appropriate for a free-expanding condensate we have neglected
possible effects of the interaction of Rb with K also during the
early phases of the expansion, which instead cannot be ruled out a
priori and need further investigation.\par In conclusion we have
produced a degenerate mixture in which a Rb BEC is fully immersed
in a K Fermi sea. The mutual attractive interaction constitutes an
important novelty, since it allows thermal contact between the FB
components to continue into the degenerate regime, where
loss-induced heating of the Fermi gas may also play a significant
role \cite{timmermans}. A deeper exploration of thermal exchange
in the degenerate regime will require the development of new
techniques for better thermometry. One possible scheme could be
based on the production of velocity-selected Rb atoms
\cite{chikkatur} for the study of their collisional relaxation in
the fermionic gas \cite{ferrari}. We also observed a first
evidence of the effect of the strong FB interaction on the density
profile of the bosons. Further insight in this phenomenology could
be obtained, for instance, through the study of the dynamics
induced by the interaction, following the release of one of the
two species from the trap \cite{vichi}.\par Finally, we note that
this K-Rb mixture opens new possibilities for the production of
ultracold heteronuclear molecules. In particular, as recently
discussed in Ref. \cite{goral}, dipolar fermions could provide a
new landscape for the quest of fermionic superfluidity.

We acknowledge stimulating discussions with G. Ferrari, who also
contributed to the initial stages of the experiment, and with M.
Modugno. We thank R. J. Brecha for a critical reading of the
manuscript. This work was supported by MIUR under a PRIN project,
by ECC under the Contract HPRICT1999-00111, and by INFM, PRA
"Photonmatter".

\vspace{2cm}


\begin{thebibliography}{99}
\bibitem{evaporazione} We use a radio-frequency knife tuned to the $|2, 2\rangle
\rightarrow |2, 1\rangle$ transition of Rb; at high temperature also K is
evaporated, but with a threshold energy larger than the one for Rb by a factor
9/4, while it is no longer affected for temperatures below 5~$\mu$K.
\end{thebibliography}
\end{document}